\documentclass{article}
\usepackage{spconf,amsmath,graphicx}
\usepackage{color} % delete this before submitting

\title{
Speech Emotion Recognition with Co-Attention based Multi-level Acoustic Information}

\name{Heqing Zou$^1$, Yuke Si$^2$, Chen Chen$^1$, Deepu Rajan$^1$, Eng Siong Chng$^1$}
\address{$^1$Nanyang Technological University, Singapore  \quad$^2$Tianjin University, China}
\begin{document}
%\ninept
%
\maketitle
\begin{abstract}
Speech Emotion Recognition (SER) aims to help the machine to understand human's subjective emotion from only audio information. However, extracting and utilizing comprehensive in-depth audio information is still a challenging task. In this paper, we propose an end-to-end speech emotion recognition system using multi-level acoustic information with a newly designed co-attention module. We firstly extract multi-level acoustic information, including MFCC, spectrogram, and the embedded high-level acoustic information with CNN, BiLSTM and wav2vec2, respectively. Then these extracted features are treated as multimodal inputs and fused by the proposed co-attention mechanism. Experiments are carried on the IEMOCAP dataset, and our model achieves competitive performance with two different speaker-independent cross-validation strategies. Our code is available on GitHub. 
\end{abstract}
\section{Introduction}
\label{sec:intro}

Automatic recognition of emotions finds several applications such as human-computer interaction \cite{cowie2001emotion} and surveillance \cite{clavel2008fear}. Some researchers propose to combine acoustic information with textual information and learn high-level context information to help make the final emotion prediction \cite{peng2021efficient}. However, the corresponding transcriptions are not always available for most emotion recognition applications. Besides, the generated text with a current automatic speech recognition (ASR) system could also introduce word recognition errors and interfere with the emotion recognition task. Emotion perception from only audio signals is much easier to implement compared with multimodal emotion recognition with additional textual and visual signals because single audio data is easier to be obtained. Transforming the speech emotion recognition (SER) problem into a multi-level fusion problem by integrating multiple acoustic information is a potentially effective method to utilize the complete audio information.

%, thus making it possible to be widely used.
The vast majority of SER problems involve extracting key audio features like Mel-frequency Cepstral Coefficient (MFCC), Constant-Q Transform (CQT) or constructing the corresponding spectrogram image to treat the problem as an image classification problem \cite{guizzo2020multi}. 
%yuke adding:
Both MFCC and spectrogram reflect more information of a speech signal in the frequency domain. MFCC can be regarded as a low-level feature based on human knowledge. Spectrogram can be further processed to obtain high-level information through a deep neural network. These methods are intuitive and simple but usually ignore time-domain information of the speech signal.

%\cite{peng2021efficient, xu2021speech}
Various encoders with different architecture details are designed for different acoustic signals, e.g., CNN for spectrogram and CNN/LSTM for MFCC. The acoustic information is mined using a series of CNNs with different kernel sizes in \cite{peng2021efficient}. Some methods propose to introduce a combination of networks to extract acoustic information, e.g., \cite{cao2021hierarchical} combine LSTM and Gated Multi-features Unit (GMU) to extract both static and dynamic speech signals. In \cite{gao2021domain}, Gao et al propose a domain-adversarial auto-encoder to extract discriminative representations with pre-trained spectrogram information. Extracting features from different sources requires the corresponding source-specific neural networks.

Different types of attention mechanisms have been proposed for processing the extracted features, like the commonly used self-attention \cite{cao2021hierarchical,wu2021emotion} and cross-modal attention \cite{sun2021multimodal}. For the models with more complex input combinations, new attention mechanisms are introduced. \cite{yoon2020attentive} fuses two modalities and then combines the result with another modality using the proposed attentive modality-hop mechanism. In \cite{li2021hierarchical}, a hierarchical attention-based temporal convolutional network is designed to fuse the inter-channel and intra-channel features for spectrogram images. 

\begin{figure*}[t]
\centering
\includegraphics[width=15cm]{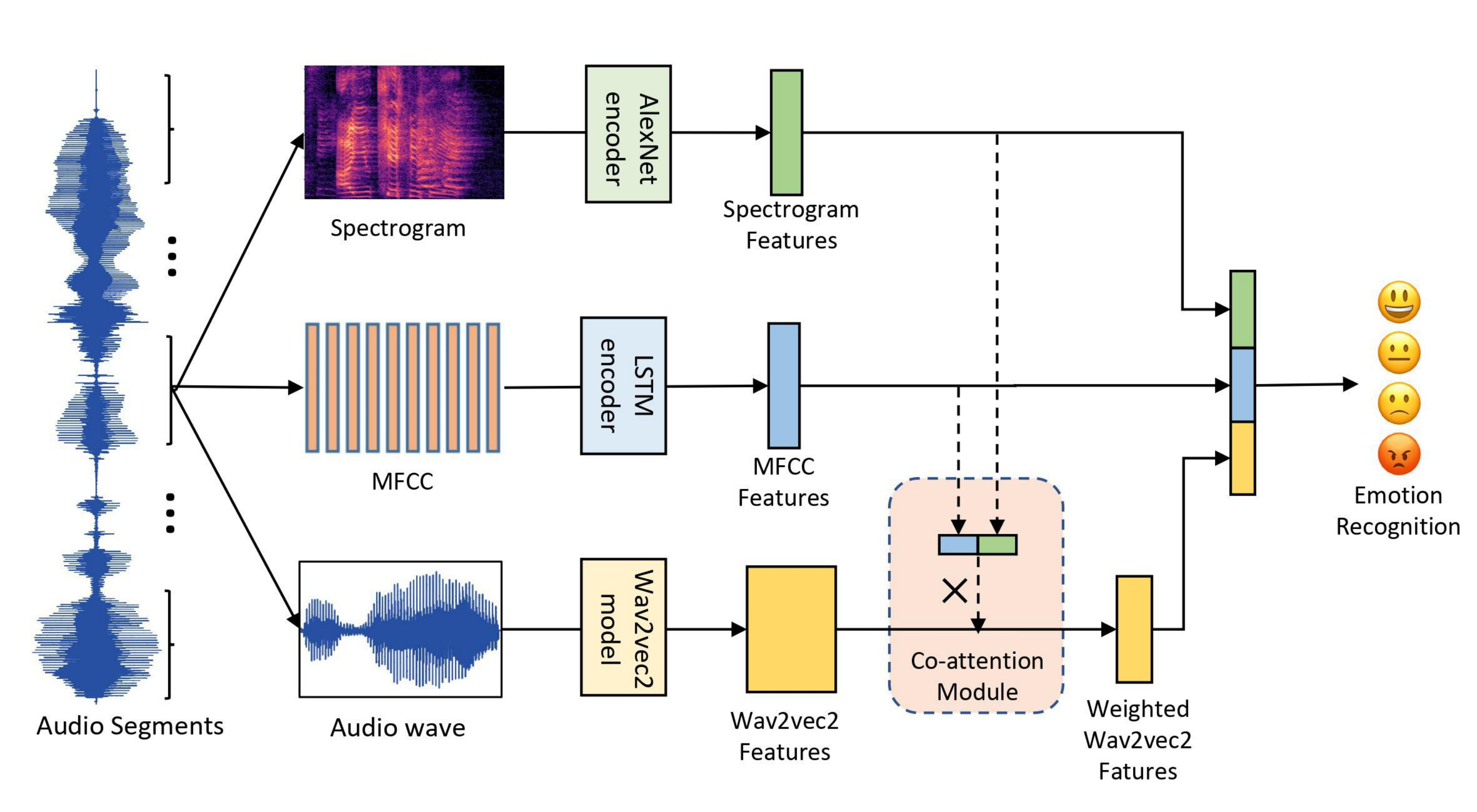}
\caption{The overall architecture of our proposed method.}
\label{fig: MainModel}
\end{figure*}

In this paper, we introduce three different encoders for multiple levels of acoustic information: CNN for spectrogram, BiLSTM for MFCC and the transformer-based acoustic extracting network wav2vec2 \cite{baevski2020wav2vec} for raw audio signals. With the designed co-attention module, we optimize to get the final wav2vec2 embedding (W2E) after weighting each frame by utilizing the effective information extracted from MFCC and spectrogram features. We concatenate all three extracted features and make the final emotion prediction with this finally fused information. The proposed model surpasses current competitive models on the widely used IEMOCAP dataset with the leave-one-speaker-out and leave-one-session-out cross-validation strategy. 

% Our main contributions can be summarized as follows:

\section{PROPOSED METHOD}
\label{sec:method}

In this section, we describe our co-attention-based SER system by integrating multiple acoustic information. Fig.\ref{fig: MainModel} shows the overall structure of our proposed method. As illustrated, after splitting the raw audio utterance into several segments, three levels of acoustic information (MFCC, spectrogram and W2E) of a segment are introduced to the respective feature encoder networks and fused with the proposed co-attention method for the final emotion recognition.

\vspace{-0.15in}

\subsection{Model Overview}
\label{ssec:overvie}

We denote the MFCC, spectrogram and wav2vec2, which are obtained from the same audio segment, as $x_m \in \mathbf{R}^{T_m\times{D_m}}$, $x_s \in \mathbf{R}^{T_s\times{D_s}}$ and $x_w \in \mathbf{R}^{T_w\times{1}}$, respectively. The extracted MFCC features $x_m'$ and spectrogram features $x_s'$ are concatenated and transformed with linear layers to get the weights for different frames of wav2vec outputs $x_w'$. After multiplication with these generated weights, we get the final W2E vector from the raw wav2vec outputs. The final obtained W2E $x_w''$ are concatenated with the previous MFCC features $x_m'$ and spectrogram features $x_s'$ for the final emotion recognition task. The generated weights of wav2vec frames from MFCC and spectrogram features and the final feature combination are denoted as $x_{coatt}'$ and $x'$, respectively. The target of the data is denoted by $y$ and the final prediction is denoted as $\hat{y}$.

\subsection{Learning with Multi-level Acoustic Information}
\label{ssec:multi-level-lerning}

Here we define the multi-level acoustic information as the combination of the human knowledge based low-level MFCC, deep learning based high-level spectrogram and W2E, thus to cover characteristics of the speech signal in both frequency and time domain. MFCC sequence is processed by a bidirectional LSTM with a dropout of 0.5 and flattened. The flattened vector is input to a linear layer with ReLU as an activation function with a dropout of 0.1 to obtain

\begin{equation}
    x_m' = f_m(BiLTSM(x_m))
\end{equation}
where $x_m' \in \mathbf{R}^{D_m'}$.

The spectrogram image is first reshaped for the pre-trained AlexNet. A similar operation as for MFCC features is conducted on the AlexNet extracted features to obtain

\begin{equation}
    x_s' = f_s(AlexNet(x_s))
\end{equation}
where $x_s' \in \mathbf{R}^{D_s'}$.

Raw audio segments are sent directly to the corresponding wav2vec2 processor and wav2vec2 model to get the target raw wav2vec2 outputs as

\begin{equation}
    x_w' = Wav2Vec2(x_w)
\end{equation}
where $x_w' \in \mathbf{R}^{T_w'\times{D_w'}}$.

\subsection{Co-attention-based Fusion}
\label{ssec:coattention}

Considering that all three acoustic information sources play a similar role in the final emotion prediction, we use the correlation among them to guide the feature adaptation. Generally, the last frame or the average of the wav2vec2 output is used to represent the wav2vec2 features. It is obvious that we lose some effective information among the sequence dimension. Here, we introduce a kind of co-attention module to combine different frames of W2E with frame weights generated by the features of MFCC and spectrogram features.

Firstly, we create a 1-dimension matrix from MFCC features $x_m'$ and spectrogram features $x_s'$ with a transformation layer given by

\begin{equation}
    x_{att}' = f_{att}(x_m' \oplus {x_s'})
\end{equation}
where $x_{att}' \in \mathbf{R}^{1\times{T_w'}}$.

The wav2vec2 outputs are multiplied with the previous generated weights to get the final weighted wav2vec2 features as

\begin{equation}
    x_{w}'' = ({x_{att}'} \cdot x_w')^{T}
\end{equation}
where $x_{w}'' \in \mathbf{R}^{{D_w'}}$.

The final MFCC, spectrogram features and the weighted W2Es are concatenated and the speech emotion prediction is written as

\begin{equation}
    \hat{y} = f(x_m' \oplus {x_s'} \oplus {x_{w}''})
\end{equation}

\subsection{Objective}
We use the commonly used cross-entropy loss for emotion classification and our objective is

\begin{equation}
    L =  L_{ce}(y-\hat{y})
\end{equation}

%The final emotion recognition decision of an utterance is depended on the average judgement of results of the sub-segments.

\label{ssec:obejctive}

\begin{table}
\centering
\caption{Performance comparison with 5-fold leave-one-session-out [12,7,13] and 10-fold leave-one-speaker-out [14,15,16,17,18,5] cross-validation strategy on IEMOCAP.}
\begin{tabular}{l|cc}
\hline
\hline
\textbf{Model} & \textbf{WA} & \textbf{UA}\\
\hline
\hline
CNN-ELM+STC attention\cite{guo2021representation} & 61.32 &  60.43 \\
Audio$_{25}$\cite{wu2021emotion} & 60.64$\pm$1.96 & 61.32$\pm$2.26 \\
IS09 - classification \cite{tarantino2019self}& 68.1 & 63.8 \\
\hline
Ours & \textbf{69.80} & \textbf{71.05}  \\

\hline
\hline
RNN(prop.)-ELM\cite{lee2015high} & 62.85 & 63.89  \\
3D ACRNN\cite{chen20183} & - & 64.74$\pm$5.44 \\
BLSTM-CTC-CA\cite{zhao2019attention} & 69.0 & 67.0  \\
CNN GRU-SeqCap\cite{wu2019speech} & 72.73 & 59.71  \\
CNN${\_}$TF${\_}$Att.pooling\cite{li2018attention} & 71.75 & 68.06  \\
HNSD\cite{cao2021hierarchical} & 70.5 & 72.5 \\
\hline
Ours & 71.64 & \textbf{72.70}  \\
\hline
\hline
\end{tabular}
\label{tab:results}
\end{table}

\section{EXPERIMENT}
\label{sec:experiment}

Our proposed method is validated on the Interactive Emotional Dyadic Motion Capture (IEMOCAP) \cite{busso2008iemocap} dataset. In this section, we firstly introduce the dataset processing and audio sources used. Then we describe our experimental setup and the used validation strategy. 

\subsection{Datasets}
\label{ssec:datasets}

IEMOCAP is a widely used emotion recognition dataset, recorded from ten different actors with audio, video, transcriptions and motion-capture information. Following others' work \cite{guo2021representation, wu2021emotion, cao2021hierarchical}, we merge “happy” and “excited” into the category of “happy” and we consider the 5531 acoustic utterances from 4 emotions, angry, sad, happy and neutral. In order to more accurately evaluate the performance of the model, we test our model with the 5-fold leave-one-session-out and the 10-fold leave-one-speaker-out cross-validation strategy to generate the speaker-independent results. Also, we use the commonly used weighted accuracy (WA) and the unweighted accuracy (UA) as the evaluation metrics.

\subsection{Experimental Setup}
\label{ssec:implementation}

The used raw audio signals are sampled at 16 kHz. We spilt each audio utterance into several segments with a length of 3 seconds. When a segment is less than 3 seconds, a padding operation with 0 will be applied to this segment to keep the same length. The final prediction result of an audio utterance will be decided by all split segments from this utterance. 

To make full use of different levels of speech information, we use three kinds of acoustic information in this SER task, MFCC, spectrogram and W2E. MFCC is a 40-dimension HTK-style Mel frequencies feature that taking into account the human auditory characteristics. It is extracted from the raw audio segments with librosa library \cite{mcfee2015librosa}. Spectrogram and the W2E are the deep features of audio signals. For spectrogram, a series of 40-ms Hamming windows with a hop length of 10 ms is applied and here we treat each windowed block as a frame. Each frame is transformed into a frequency domain with the Discrete Fourier Transform (DFT) of length 800. The first 200 DFT points are used as input spectrogram features. We finally get a spectrogram image with a size of 300*200 for each audio segment. Like the multimodal emotion recognition method \cite{cai2021speech}, W2E are obtained from the pre-trained transformer-based wav2vec2 network. It is the reflection of the deep feature of speech in the time domain.

This SER system is implemented in PyTorch. The optimizer for the model is AdamW with a learning rate of 1e-5. The training batch size is 64 and we set the early stopping setting as 8 epochs. Our code will be available on Github\footnote{https://github.com/Vincent-ZHQ/CA-MSER}.

\section{Results and Analysis}
\label{sec:results}

In this section, we present the model performance and design an ablation study to evaluate the influence of different inputs and used modules. We also visualize the extracted features of our model with t-distributed stochastic neighbour embedding (t-SNE) and the final normalized confusion matrix.

\begin{table}
\centering
\caption{Ablation study on the proposed model.}
\begin{tabular}{c|cc}
\hline
\hline
\textbf{Model} & \textbf{WA} & \textbf{UA}\\
\hline
MFCC & 57.60 & 58.09 \\
Spectrogram & 62.13 & 62.25 \\
W2E & 64.03  & 65.67 \\
\hline
%MFCC+Spectrogram & 62.44 & 63.04  \\
MFCC+W2E (w/o co-att) & 64.62 & 65.93  \\
Spectrogram+W2E (w/o co-att) & 66.20 & 67.22 \\
MFCC+Spectrogram+W2E (w/o co-att) & 67.22 & 67.81  \\
\hline
W2E (w/ co-att) & 67.55 & 68.65  \\
MFCC+W2E (w/ co-att) & 69.11 & 70.30  \\
Spectrogram+W2E (w/ co-att) & 70.05 & 71.30  \\
MFCC+Spectrogram+W2E (w/ co-att) & \textbf{71.64} & \textbf{72.70}  \\
\hline
\hline
\end{tabular}
\label{tab:ablation01}
\end{table}

\subsection{Results and Comparison}
\label{ssec:resultscom}

As shown in Table \ref{tab:results}, our proposed method could achieve the best performance of $69.80\%$ and $71.05\%$ in terms of UA and WA for the leave-one-session-out validation strategy. And for the leave-one-speaker-out validation strategy, this method could also achieve the highest UA with a value of $72.70\%$. At the same time, its performance in WA is also competitive with a very similar result of $71.64\%$ compared with UA on this unbalanced IEMOCAP dataset.

\subsection{Ablation Study}
\label{ssec:ablationstudy}
Our proposed method utilizes the multiple levels of acoustic information, which contains the time domain and frequency domain. Table \ref{tab:ablation01} shows the ablation study of model performance with different combinations of acoustic information. The first three rows are the emotion recognition results with only one level of acoustic information: MFCC, spectrogram and W2E. W2E provides better performance than the others for the final emotion recognition. The next three rows summarize the results from the combination of different features with W2E. The last four rows present the results of different combination features with the weighted W2E information after co-attention. The combination of multiple acoustic information and the proposed co-attention module are observed to contribute a lot to improve the whole model's performance. 

\begin{figure}[htb]
\begin{minipage}[a]{.48\linewidth}
  \centering
  \centerline{\includegraphics[width=4.5cm]{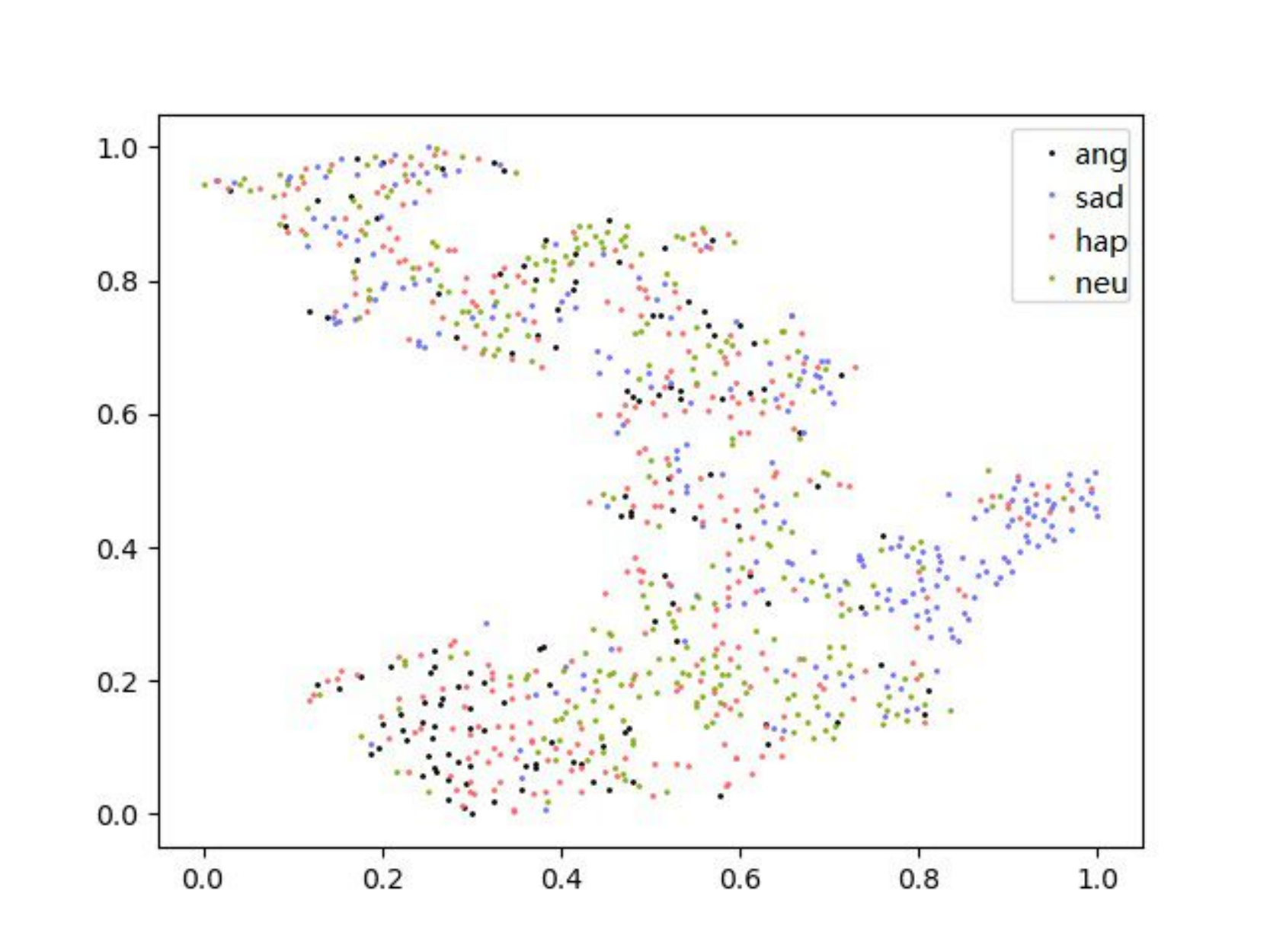}}
%  \vspace{1.5cm}
  \centerline{(a)W2E w/o co-att
}\medskip
\end{minipage}
\hfill
\begin{minipage}[a]{0.48\linewidth}
  \centering
  \centerline{\includegraphics[width=4.5cm]{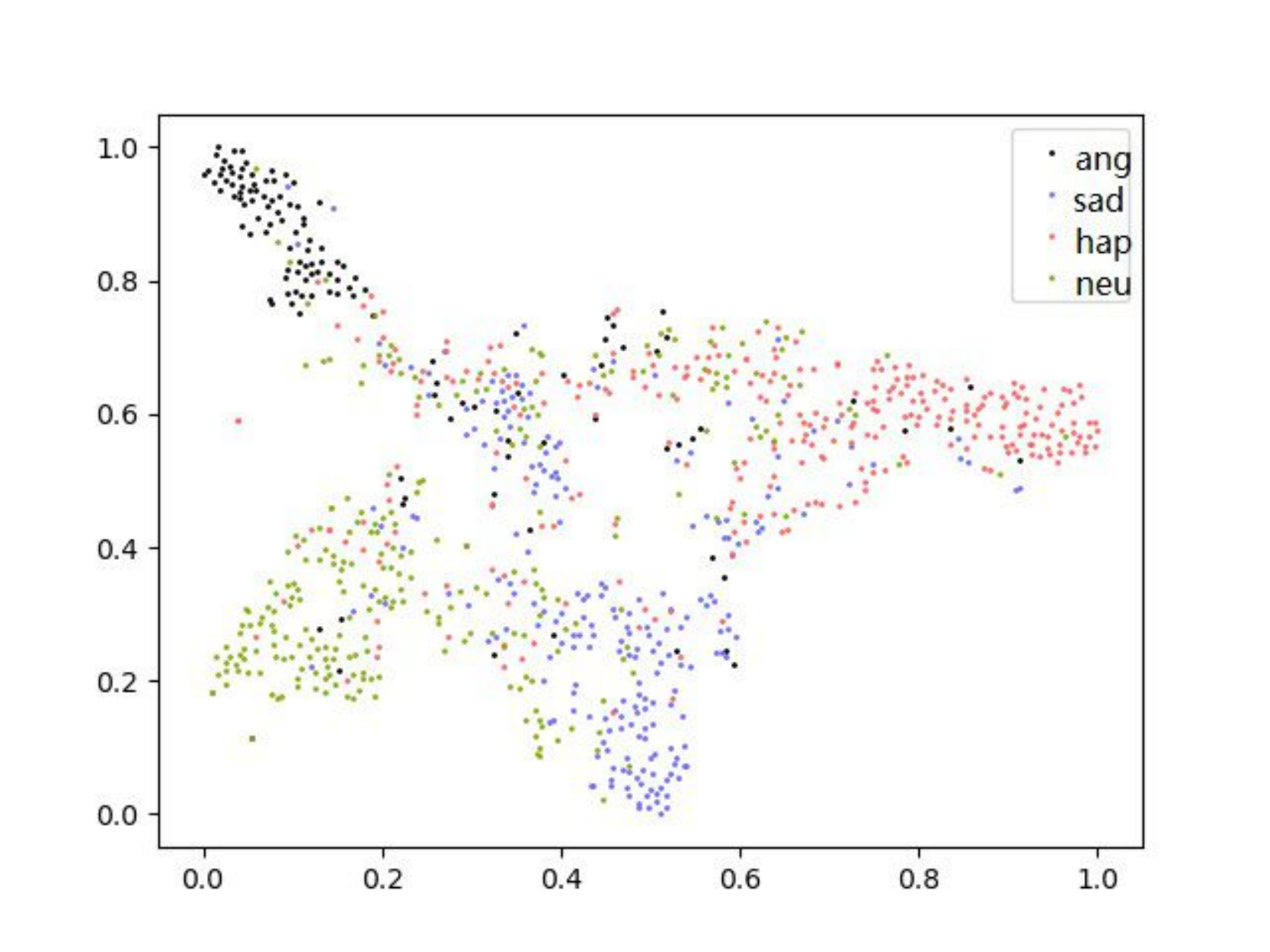}}
%  \vspace{1.5cm}
  \centerline{(b)W2E w/ co-att
}\medskip
\end{minipage}
\begin{minipage}[c]{.48\linewidth}
  \centering
  \centerline{\includegraphics[width=4.5cm]{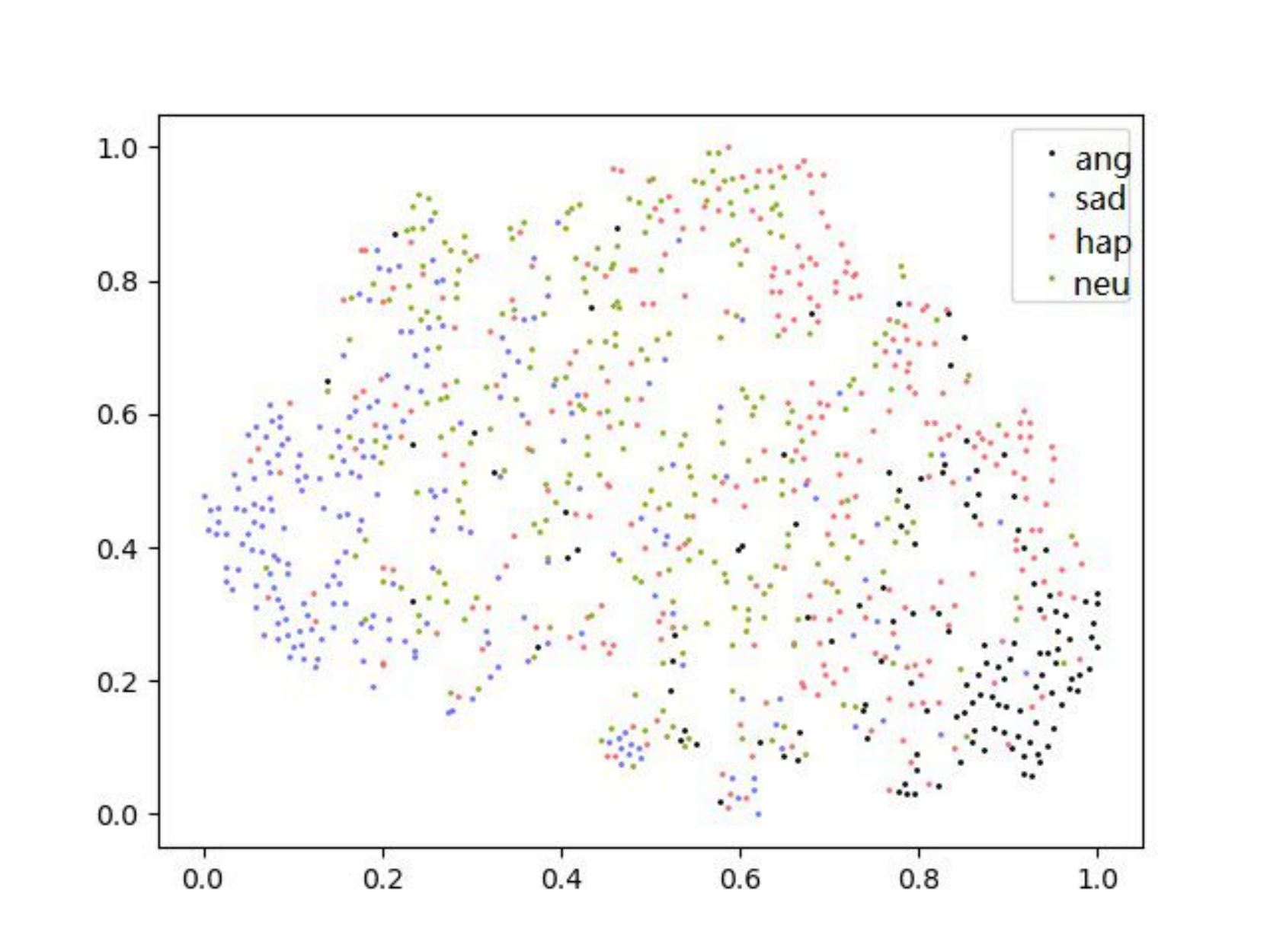}}
%  \vspace{1.5cm}
  \centerline{(c)Final features w/o co-att
}\medskip
\end{minipage}
\hfill
\begin{minipage}[c]{0.48\linewidth}
  \centering
  \centerline{\includegraphics[width=4.5cm]{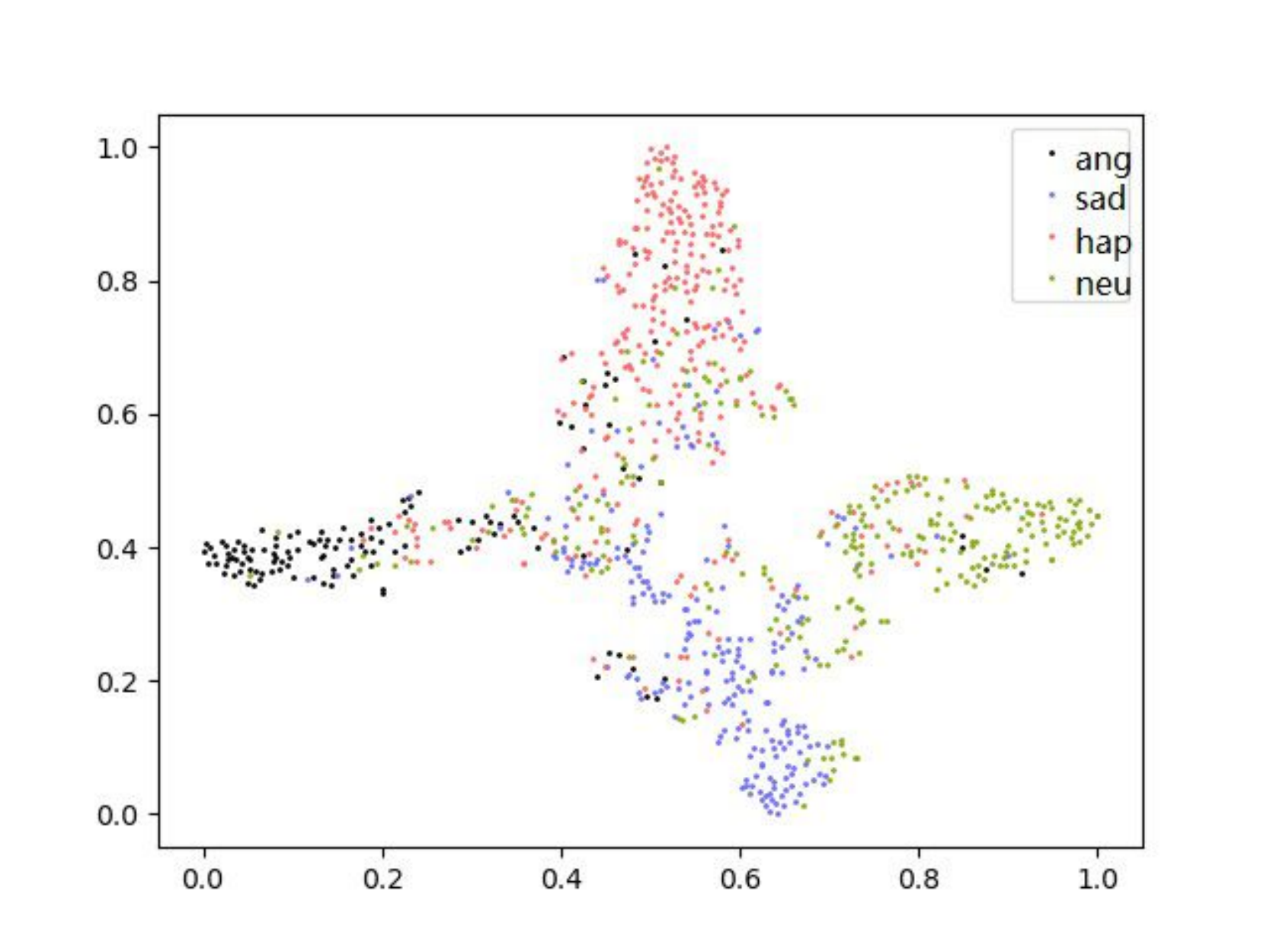}}
%  \vspace{1.5cm}
  \centerline{(d)Final features w/ co-att
}\medskip
\end{minipage}
\vspace{-0.15in}
\caption{The t-SNE visualization of feature distribution. (a) and (b) are the final extracted W2Es in the model trained with multi-level acoustic information without and with the proposed co-attention. (c) and (d) are the final combined features without and with the proposed co-attention}
\label{fig:tsne}
\end{figure}

\begin{figure}[htb]
\begin{minipage}[a]{.48\linewidth}
  \centering
  \centerline{\includegraphics[width=4.0cm]{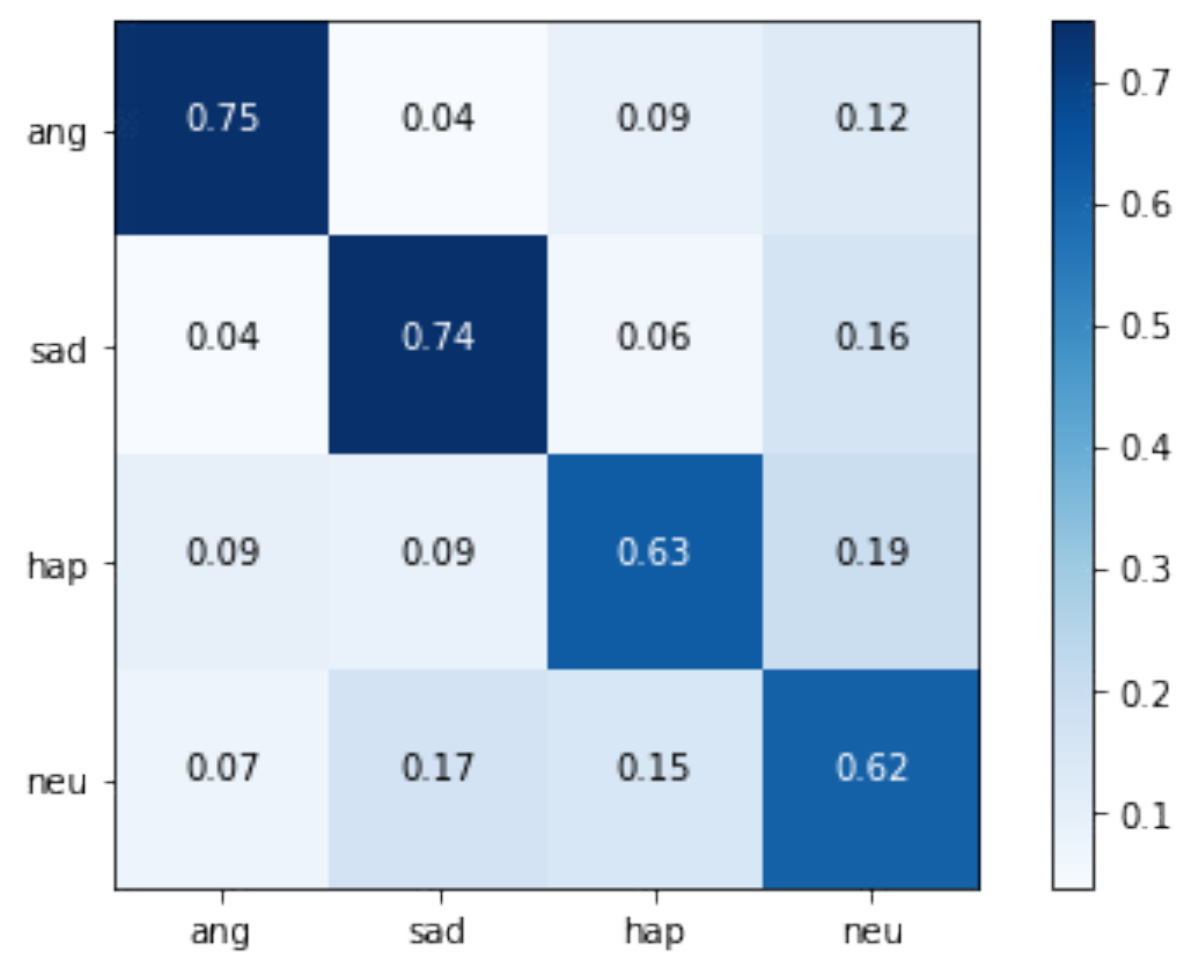}}
%  \vspace{1.5cm}
  \centerline{(a) w/o co-att
}\medskip
\end{minipage}
\hfill
\begin{minipage}[a]{0.48\linewidth}
  \centering
  \centerline{\includegraphics[width=4.0cm]{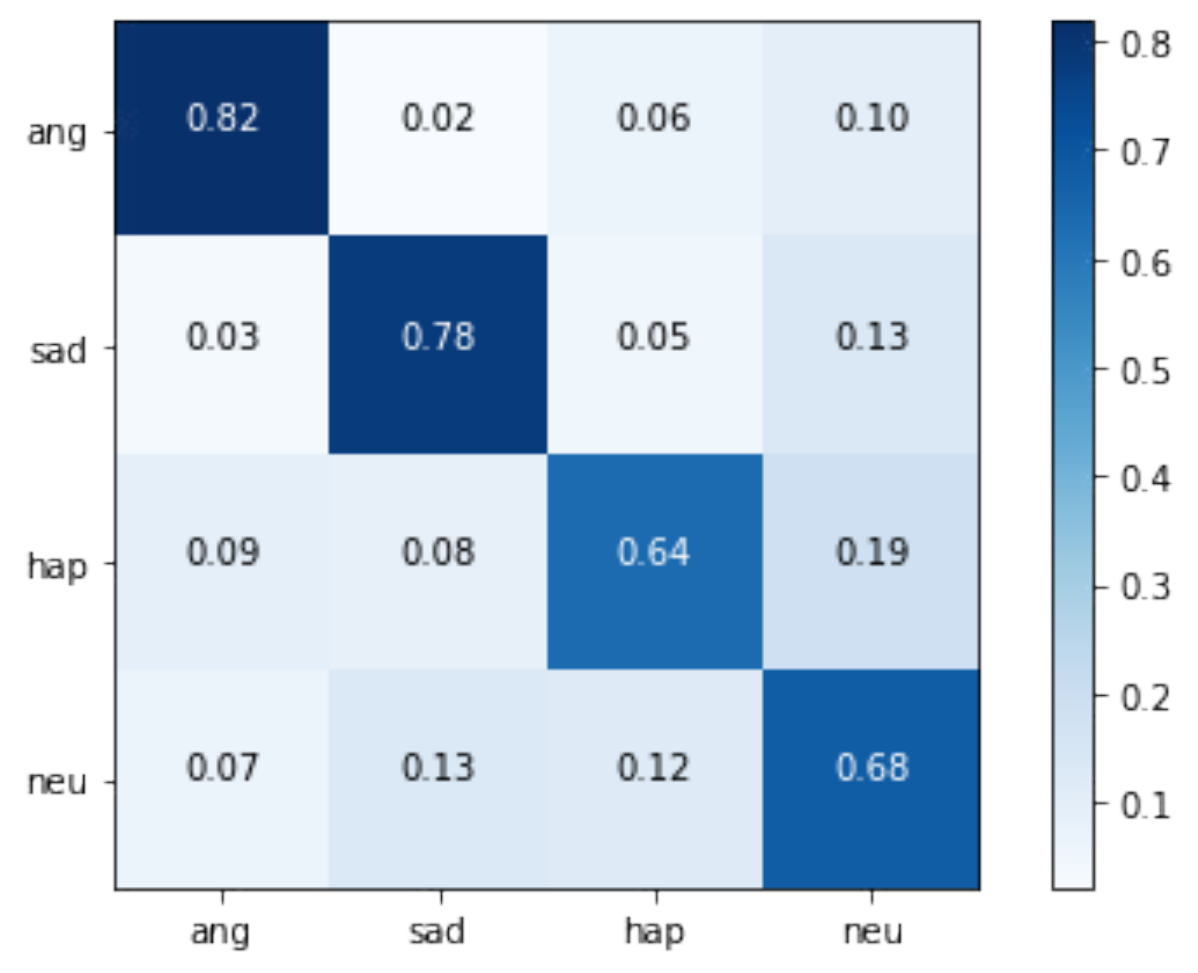}}
%  \vspace{1.5cm}
  \centerline{(b) w/ co-att
}\medskip
\end{minipage}
\vspace{-0.15in}
\caption{The normalized confusion matrix for the final speech emotion recognition without and with the proposed co-attention module.}
\label{fig:cm}
\end{figure}

The ablation study also shows the effectiveness of the proposed co-attention mechanism. From the last two rows of Table \ref{tab:ablation01}, the co-attention mechanism further optimizes the fused data and performs better than the direct concatenating operation with $4.42\%$ and $4.89\%$ improvement on WA and UA, respectively. As shown in Fig.\ref{fig:tsne}, the t-SNE visualization of the weighted W2E and final combined features after co-attention present a much more clear classification boundary when compared with the results of the unweighted W2E and final combined features without co-attention. From Fig. \ref{fig:cm}, we also observe that the final classification results of the model with co-attention are much better than the model without co-attention from the final normalized confusion matrix. 

\section{Conclusion}
\label{sec:conslusion}

This paper proposes a co-attention-based SER system utilizing multi-level acoustic information. By designing different encoders, this model could get feature-specific information from the raw audio signals and enables complementary acoustic information for the SER problem. Also, this method introduces a co-attention based fusion method for getting weighted wav2vec2 embeddings and combining the final features. The experiments on the IEMOCAP dataset show that our proposed method achieves competitive performance with different speaker-independent cross-validation methods. In the future, we would like to combine the knowledge from different languages or datasets to improve the final performance.

%\vfill
%\pagebreak

\vfill\pagebreak

%\bibliographystyle{IEEEbib}
\bibliographystyle{IEEEtran}
\bibliography{refs}

\end{document}